\documentclass[twocolumn,preprintnumbers,amsmath,amssymb]{revtex4}
\usepackage{epsfig}
\usepackage{dcolumn}% Align table columns on decimal point
\usepackage{bm}% bold math
\usepackage{xspace}
\newcommand{\as}{\ensuremath{\alpha_s}\xspace}
\begin{document}

\preprint{CERN-PH-TH/2008-036, NIKHEF/2008-002}

\title{A New Framework for Multijet Predictions and its application to Higgs
  Boson production at the LHC}

\author{Jeppe R.~Andersen}
\affiliation{Theory Division, Physics Department, CERN, CH 1211 Geneva 23, Switzerland}

\author{Chris D.~White}
\affiliation{Nikhef, Kruislaan 409, 1098 SJ Amsterdam, The Netherlands}

\begin{abstract}
  We present a new framework for calculating multi-jet observables through
  resummation. The framework is based on the factorisation of scattering
  amplitudes in an asymptotic limit. By imposing simple constraints on the
  analytic behaviour of the result when applied away from this limit, we get
  good agreement with the known lowest order perturbative behaviour of the
  scattering amplitude, and predictions for the behaviour to all orders in
  the perturbative series. As an example of application we study predictions
  for Higgs Boson production through gluon fusion at the LHC in association
  with at least two jets.
\end{abstract}

\pacs{Valid PACS appear here}% PACS, the Physics and Astronomy

\maketitle

Events with many energetic particles will form the backbone of many search
strategies at the LHC for physics beyond the Standard Model (SM) of particle
phy\-sics, but will simultaneously test our understanding of SM
processes. Energetic particles of QCD charge will be detected as jets, and
events with multiple jets require the calculation of scattering amplitudes
with a high number of external quark and gluon (parton) legs. However, the
calculation of the SM contribution to such processes beyond even the lowest
order in perturbation theory is notoriously difficult. Despite recent
impressive progress in the calculation of many multi-leg scattering
amplitudes at both tree and one-loop level, the first two radiative
corrections in the perturbative series (i.e.~full next-to-next-to-leading
accuracy) for observables at the LHC are known only for very few
cases\cite{Harlander:2002wh,Anastasiou:2002yz,Anastasiou:2003ds}.

One estimate of the effect of higher order corrections beyond what is
currently calculable in full fixed-order perturbation theory can be obtained
by interfacing the fixed perturbative calculation with \emph{parton shower}
programs\cite{Sjostrand:2007gs,Bahr:2008pv}, which sum the effect of further
\emph{soft and collinear} radiation (i.e.~of low invariant mass)
to all orders in perturbation theory. % This is achieved by utilising the
% factorisation of QCD soft and collinear radiation.

As we will demonstrate in this paper, there are important processes where the
perturbative corrections from \emph{hard} radiation (i.e.~of large invariant
mass) is sizeable. In such cases, the parton shower approach cannot be
expected to capture the dominant effect of higher order corrections. In this
letter we will develop an alternative formalism for estimating the all-order
perturbative corrections, which does not rely on the description of only soft
and collinear radiation. Instead, the formalism will utilise results on the
factorisation of matrix elements in the opposite limit of infinite invariant
mass between all radiated particles. We will see that without further
modifications, the results obtained in this limit lead to a poor approximation
of LHC scattering amplitudes. However, known analytic properties of the full
perturbative amplitude can be
incorporated, % leading to a good agreement with
% the results from full fixed order perturbation theory, where these are known,
% while allowing for the calculation to be extended to all orders in
% perturbation theory.
%
such that the approximate amplitudes agree well with the full result at low
orders in perturbation theory. The approximate amplitudes can then be applied
to any order in the perturbation expansion, where the full results are
incalculable.

We will validate the approach by considering Higgs boson production through
gluon-gluon fusion (GGF) (mediated by a top-quark loop) in association with
at least two jets (the $hjj$-channel). We will explicitly compare our
approximate amplitudes with those obtained using full fixed order
perturbation theory, where such results are known. This process can be used
to search for the Higgs boson~\cite{Klamke:2007cu}, and also potentially to
measure its couplings to the top quark~\cite{Hankele:2006ja}. The
$hjj$-channel also has a large contribution from \emph{weak boson fusion}
(WBF)\cite{Cahn:1983ip,Dicus:1985zg,Altarelli:1987ue}, and by
suppressing the GGF-contribution it is possible to measure the
coupling of any Higgs boson candidate to the electro-weak bosons, and thus
determine\cite{Plehn:2001nj} whether the properties of the 
candidate match those of the SM Higgs boson. The suppression of the
GGF-contribution
 is achieved\cite{DelDuca:2001fn} to some degree by
applying specific event selection criteria, as discussed in
Table~\ref{tab:cuts}. However, only by calculating higher order corrections
is it possible to estimate the efficiency of such cuts.

The current state of the art in fixed order perturbation theory for Higgs
boson production through GGF in association with at least two jets includes
only the first perturbative corrections, and is presented in
Ref.~\cite{Campbell:2006xx}. The effects of further soft and collinear
radiation were studied in Ref.~\cite{DelDuca:2006hk}. In the limit of
infinite top mass, the coupling of the Higgs boson to gluons through a top
quark loop can be described as a point
interaction\cite{Wilczek:1977zn,Dawson:1990zj,Djouadi:1991tk}. This
approximation was applied in all these studies, and will be applied also in
the present one, although this is not essential to the approach.

We will start motivating the need for considering the hard radiative
corrections to higher orders by studying the leading order predictions in
full QCD for Higgs boson production in association with hard jets.  We apply
similar event selection cuts to the study of Ref.\cite{Campbell:2006xx} as
detailed in table~\ref{tab:cuts}\footnote{If further a rapidity veto on jets
  is applied, the effects studied in Ref.~\cite{Forshaw:2007vb} may need to be taken into
  account.}. All of the following results are obtained by choosing
renormalisation and factorisation scales in accordance with the study of
Ref.\cite{DelDuca:2006hk}, and the following values for the Higgs boson mass,
vacuum expectation value of the Higgs boson field and top quark mass
respectively:
\begin{equation}
m_H=120\text{GeV}, \langle \phi\rangle_0=\frac v {\sqrt 2}, v=246\text{GeV}, m_t=174\text{GeV}.\nonumber
%\label{params}
\end{equation}
We also include a factor multiplying the effective Higgs Boson vertices,
accounting for finite top-mass effects~\cite{Dawson:1993qf}:
\begin{equation}
K(\tau)=1+\frac{7\tau}{30}+\frac{2\tau^2}{21}+\frac{26\tau^3}{525},\quad \tau=\frac{m_H^2}{4m_t^2}.\nonumber
%\label{K}
\end{equation}
We choose the $k_t$-jet algorithm as implemented in
Ref.\cite{Butterworth:2002xg} with $R=0.6$, and the parton distribution
functions of Ref.~\cite{Martin:2004ir}. For the strong coupling \as, we
choose renormalisation scales as in Ref.~\cite{DelDuca:2006hk} (\footnote{For
  the resummed results presented later, some freezing of the coupling
  \as is necessary below a suitable low scale $Q_0$. However, the
  results are fairly insensitive to this choice.}).
\begin{table}[tbp]
  \centering
  \begin{tabular}{|rl||rl|}
    \hline
    $p_{j_\perp}$ & $> 40$ GeV & $y_c\cdot y_d$& $<0$ \\
    $|y_{j,h}|$ & $<$ 4.5 & $\vert y_c-y_d \vert$ & $> 4.2$ \\\hline
  \end{tabular}
  \caption{The cuts used for all results in this letter, in terms of rapidities $y_i$
 and transverse momenta $p_{i,\perp}$. The suffices $c,d$ label
cuts that must be satisfied by at least two jets, whereas $j$ labels conditions that must
be satisfied by all jets; $h$ labels the Higgs boson.}
  \label{tab:cuts}
\end{table}
With this, we find (using matrix
elements from \texttt{MADEvent/MADGraph}\cite{Alwall:2007st}) the tree-level
cross-section from the QCD generated $hjj$-channel to be
$281^{+210}_{-111}$fb, where the uncertainty is obtained by varying the
common factorisation and renormalisation scale by a factor of two. For the
three jet sample, all jets must satisfy the left-most cuts of
Table~\ref{tab:cuts}, but we also require that there exist two jets $c,d$ satisfying all
the right-most cuts in Table~\ref{tab:cuts}. We then find the leading order
cross section for the production of Higgs Boson plus three jets ($hjjj$) to be
$257^{+262}_{-120}$fb. The requirement of an extra hard jet with an
accompanying \as-suppression leads only to a 9\% suppression compared to the
leading order prediction for $hjj$. The \as-suppression of the matrix
element is compensated by the integration over a large phase space for the
third jet. The large size of the three-jet rate (which obviously depends on
the event selection cuts) was already reported in
Ref.\cite{DelDuca:2004wt}, and clearly demonstrates the necessity of
considering hard multi-parton emissions.

We will now describe a method for approximating the perturbative scattering
matrix elements for multi-particle production to any order. We start by
recalling the result of Fadin and collaborators (FKL)\cite{Fadin:1975cb},
that in the limit of infinite invariant mass between all produced particles
(the Multi-Regge-Kinematic (MRK) limit), the leading contribution is given by
processes of the form
\begin{align}
  \alpha(p_a) + \beta(p_b) \to \alpha(p_0) + \sum_{i}^{n-1} g(p_i) + \beta(p_n)+h(p_h)\nonumber
\end{align}
where $\alpha,\beta\in\{q,\bar q,g\}$, and the partons are ordered according
to rapidity in both the initial and final state. These processes allow
neighbouring particles to be connected by gluon propagators of momentum
$q_i$, such that $p_i=q_i-q_{i+1}$. We have explicitly checked that at leading order, partonic
configurations which are not captured in this framework account for only
0.8fb (0.3\%) and 24fb ($<10\%$) of the two and three-jet rate respectively,
even when there is no requirement of large invariant mass between all
particles.

In the MRK limit, the scattering amplitude for the remaining configurations
factorises, and the results of Ref.\cite{Fadin:1975cb} can straightforwardly
be modified to include also the production of a Higgs boson. In the case of
the Higgs boson being produced between the jets (in rapidity) these
amplitudes take the form:
\begin{small}
  \begin{align}
    \begin{split}
      i&{\cal M}_{\mathrm{FKL}}^{p_ap_b\rightarrow p_0\ldots
        p_jhp_{j+1}p_n}=2i\hat s
      \left(i g_s f^{ad_0c_1} g_{\mu_a\mu_0}\right)\\
      &\cdot
      \prod_{i=1}^j \left(\frac{1}{q_i^2}\exp[\hat\alpha(q_i)(y_{i-1}-y_i)]\left(i g_s f^{c_id_ic_{i+1}}\right)C_{\mu_i}(q_i,q_{i+1})\right)\\
      &\cdot\left(\frac 1
        {q_h^2}\exp[\hat\alpha(q_i)(y_{j}-y_h)]C_{H}(q_{j+1},q_{h})\right)\label{FKL}
      \\
      &\cdot\prod_{i=j+1}^n
      \left(\frac{1}{q_i^2}\exp[\hat\alpha(q_i)(y'_{i-1}-y'_i)]\left(i g_s f^{c_id_ic_{i+1}}\right)C_{\mu_i}(q_i,q_{i+1})\right)\\
      &\cdot\frac 1 {q_{n+1}^2}\exp[\hat\alpha(q_{n+1})(y'_{n}-y_b)]\left(i
        g_s f^{bd_{n+1}c_{n+1}} g_{\mu_b\mu_{n+1}}\right),
    \end{split}
  \end{align}
\end{small}
\noindent where $g_s$ is the strong coupling constant ($\alpha_s=\frac{g_s^2}{4\pi}$);
$f^{abc}$ colour structure constants; $y_i,y_i'$ are the rapidities of the
emitted particles; $\hat{s}=(p_a+p_b)^2$ is the partonic centre of 
mass energy. The factor $C_{\mu_i}$
is a {\it Lipatov effective vertex} describing the emission of gluon $i$. This has the explicit form:
\begin{footnotesize}
  \begin{align}
    \!\!\!\!\!\!\!\!C^{\mu_i}(q_i,q_{i+1})=\left[(q_i+q_{i+1})_\perp^{\mu_i}-\left(\frac{{\hat{s}}_{ai}}{\hat{s}}+2\frac{q_{i+1}^2}{\hat{s}_{bi}}\right)p_b^{\mu_i}+\left(\frac{\hat{s}_{bi}}{\hat{s}}+2\frac{\hat{q}_i^2}{\hat{s}_{ai}}\right)p_a^{\mu_i}\right],
    \label{eq:lip1}
  \end{align}
\end{footnotesize}
where $\hat{s}_{ai}=2p_a\cdot p_i$ and similarly for $\hat{s}_{bi}$. The notation $q_\perp$ denotes the 
projection of a 4-momentum onto its transverse components.
Also, $C_H$ is an effective vertex coupling the Higgs to off-shell gluons via a top quark loop, whose
form has been calculated in~\cite{DelDuca:2003ba}.
The exponential factors $\hat{\alpha}(q_i)$
encode the leading virtual corrections (see e.g.~Ref~\cite{Fadin:1998sh}):
\begin{equation}
  \hat
  \alpha(q_i)=-\frac{g_s^2\ N_c \ \Gamma(1-\varepsilon)}{(4\pi)^{2+\varepsilon}}\frac
  2 \varepsilon\left(|q_{\perp i}|^2/\mu^2\right)^\varepsilon,
\label{alpha}
\end{equation}
where singularities have been regularised using dimensional
regularisation in $D=4+2\varepsilon$ dimensions, where $\mu$ is the
renormalisation scale and $N_c$ the number of colours. The colour factors in
equation (\ref{FKL}) are for incoming gluons. The form of the amplitude is
the same for initial state quarks, apart from different colour factors.

Eq.~(\ref{FKL}) formally applies in the so-called {\it Multi-Regge-kinematic}
(MRK) limit. Thus, it is clear why this may be a good starting point for
describing matrix elements with many hard partons.  Firstly, it can be
applied at any order in the perturbation expansion. Secondly, it does not
rely upon soft and collinear approximations.

The multi-gluon emissions have in fact not previously been studied directly
as implemented in Eq.~(\ref{FKL}). Instead, a simplified version has been
used, which is equivalent in the MRK limit. In this limit, the squared
4-momenta fulfil $q_i^2\to-|q_{\perp i}|^2$, and the squared Lipatov vertices
$-C_{\mu_i} C^{\mu_i}\to4\frac{|q_{\perp i}|^2|q_{\perp i+1}|^2}{|k_{\perp
    i}|^2}$. This means that in the products of Eq.~(\ref{FKL}), only squares
of \emph{transverse} momenta appear.  Extending these kinematic
approximations to all of phase space (not just the MRK limit), the sum over
$j,n$ and the phase space integral over the emitted gluons can be
approximated by solving the \emph{BFKL-equation}\cite{Balitsky:1978ic}. In
this form, the framework has previously been extensively applied to other
processes. In the present context (after implementing local 4-momentum
conservation, which is strictly sub-leading in the BFKL approach), we find by
expanding the solution in powers of \as that the lowest order BFKL results
for the $hjj$ ($554$fb) and $hjjj$ ($775$fb) cross sections differ from their
full leading order counterparts by $97\%$ and $200\%$ respectively. The
kinematic approximations are clearly inadequate in describing amplitudes in
general at the LHC.

Instead, we define a set of amplitudes based on Eq.~(\ref{FKL}) as written,
supplemented by the following guidelines:
\begin{enumerate}
\item Use of full virtual 4-momenta: Rather than substituting
  $q_i^2\to-|q_{i\perp}|^2$ as in the BFKL equation, we keep the dependence on
  the full 4-momenta of all particles. This ensures that outside of the MRK
  limit, the singularity structure of the approximate amplitudes coincides
  with known singularities of the full fixed order scattering amplitude.
\item Positivity of the squared Lipatov vertex: The square of the amplitudes in
  Eq.~(\ref{FKL}) are not positive definite, when the effective Lipatov
  vertex is applied to momentum configurations very far from the MRK
  limit. It is here possible to obtain $-C_{\mu_i}C^{\mu_i}<0$, where the minus
  sign arises from the contraction of the gluon polarisation tensor. We choose to
  remove the contribution from the \emph{small} region of phase space where
  this happens. 
  \end{enumerate}
These modifications combine known analytic behaviour of the full scattering
amplitudes and the factorised expressions obtained in the MRK limit to any
order in perturbation theory.

\begin{figure}
\begin{center}
\epsfig{width=\columnwidth,file=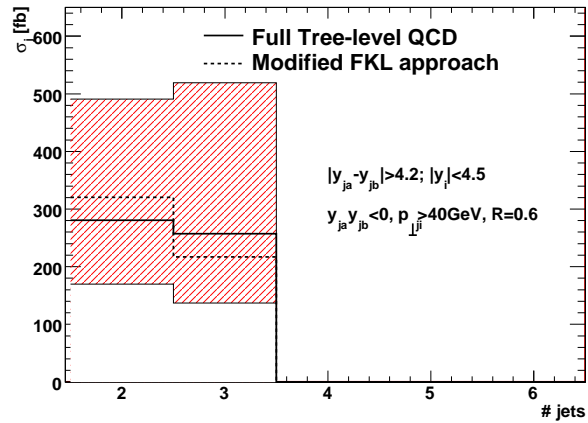}
%\scalebox{0.4}{\includegraphics{H_NJets_fixed.eps}}
\caption{The 2 and 3 parton cross-sections calculated using the known LO
  matrix elements (solid), and the estimate gained from the modified high
  energy limit (dashed). The uncertainty band on the LO result corresponds to
 scale variation by a factor of two.}
\label{LO}
\end{center}
\end{figure}

When these modifications are made, the resulting amplitudes do indeed approximate
well the known perturbative results at low orders, and thus can be reliably used at higher 
orders, where full results are unknown or computationally unfeasible. Using our approach,
we find an $\as^4$ contribution to $hjj$ of 321fb and
$\as^5$-contribution to $hjjj$ of 217fb, within 16\% and 7\%
of the full results respectively. In general, we find good agreement in a large region of phase
space, and the level of accuracy reported here does not require any fine
tuning of cuts. This is summarised in Fig.~\ref{LO}, and one sees that the approximate results
are well within the uncertainty associated with the full results, obtained by varying a common 
renormalisation and factorisation scale by a factor of two.
\begin{figure}[tb]
\begin{center}
\epsfig{width=\columnwidth,file=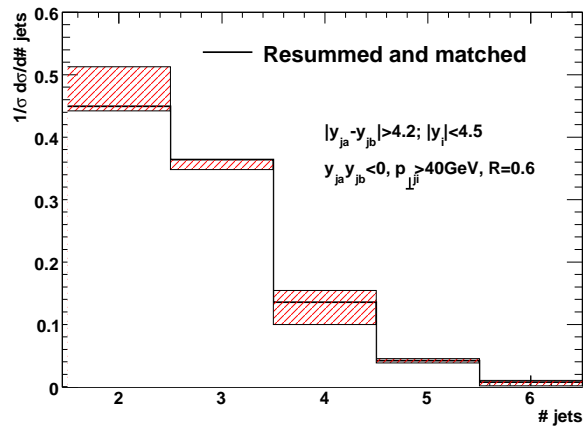}
%\scalebox{0.4}{\includegraphics{H_NJets_fixed.eps}}
\caption{Distribution of number of jets, obtained using the modified FKL approach, matched to the exact tree level results for 
2 and 3 partons. The shaded uncertainty bands are obtained by scale variations.}
\label{fig:FKL}
\end{center}
\end{figure}

Having validated the approximation at low orders in \as (where it can be
compared with known fixed order results), we now consider results obtained
using matrix elements with any number of final state partons, which at
present cannot be calculated using standard perturbation theory.
The divergence in Eq.~(\ref{FKL}) arising 
when any $p_i\to 0$ is regulated by
the divergence of the virtual corrections encoded in $\hat\alpha$. 
Thus the resulting formalism is efficiently
implemented in a Monte Carlo generator following the method for phase space generation outlined in
Ref.~\cite{Andersen:2006sp}. Given that one knows the full tree level results
for 2 and 3 partons, however (and they are also computationally quick), we have combined 
these results with the approximate matrix elements using a suitable matching procedure.
We find a total cross section of $499^{+527}_{-307}$fb. The large uncertainty
in the total cross section due to scale variations however is not reflected
in the distribution of the number of hard jets as
shown in Fig.~\ref{fig:FKL}. One sees a significant number of events with
more than 3 hard jets.

\begin{figure}[tb]
\begin{center}
\epsfig{width=\columnwidth,file=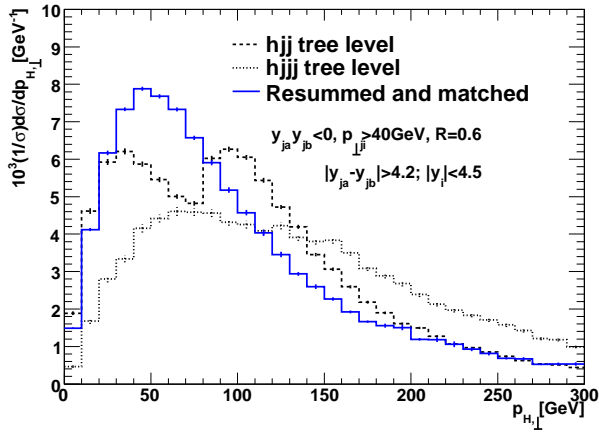}
%\scalebox{0.4}{\includegraphics{H_ptH.eps}}
\caption{The transverse momentum spectrum of the Higgs boson in the process
  $pp\to h+n$jets at tree-level for $n=2$, $n=3$, and the transverse momentum
  spectrum obtained in the resummed approach presented in this paper.}
\label{fig:ptH}
\end{center}
\end{figure}
The transverse momentum spectrum of the Higgs boson when produced in
association with at least two hard jets is shown in Fig.~\ref{fig:ptH}. To
the best of our knowledge, this is the first report of this quantity, in
contrast to the completely inclusive Higgs boson $p_t$ spectrum, which has
previously been studied in the literature.  The tree level 2 parton final
state predicts a bimodal structure, which ultimately arises from the
azimuthal correlation between the jets. This structure disappears when extra
radiation is added, giving a qualitatively different behaviour. The
significant difference between the fixed-order spectra emphasises the
importance of considering yet higher order corrections.

In summary, we have outlined a technique, not relying on a soft and collinear
approximation, for estimating scattering amplitudes with multiple partons to
all orders in perturbation theory, and demonstrated its application to Higgs boson production (via
GGF) in association with at least two jets. Our technique is based on the FKL
factorisation formula of Ref.~\cite{Fadin:1975cb}, with important
modifications which ensure that the singularity structure of the amplitudes
coincides with known all-order analytic properties of the perturbation
expansion. At low orders in \as, where the full fixed order result can be
obtained, our description agrees well, which verifies the trustworthiness of
the approach. It captures both real and virtual corrections, and can be
applied at any order in \as in a computationally efficient manner.

\section*{ACKNOWLEDGEMENTS}
CDW is funded by the Dutch Organisation for Fundamental Matter Research
(FOM). We thank Vittorio Del Duca, Eric Laenen, Gavin Salam and Jos
Vermaseren for encouraging conversations, and also the Galileo Galilei
Institute for Theoretical Physics.


\begin{thebibliography}{28}
\expandafter\ifx\csname natexlab\endcsname\relax\def\natexlab#1{#1}\fi
\expandafter\ifx\csname bibnamefont\endcsname\relax
  \def\bibnamefont#1{#1}\fi
\expandafter\ifx\csname bibfnamefont\endcsname\relax
  \def\bibfnamefont#1{#1}\fi
\expandafter\ifx\csname citenamefont\endcsname\relax
  \def\citenamefont#1{#1}\fi
\expandafter\ifx\csname url\endcsname\relax
  \def\url#1{\texttt{#1}}\fi
\expandafter\ifx\csname urlprefix\endcsname\relax\def\urlprefix{URL }\fi
\providecommand{\bibinfo}[2]{#2}
\providecommand{\eprint}[2][]{\url{#2}}

\bibitem[{\citenamefont{Harlander and Kilgore}(2002)}]{Harlander:2002wh}
\bibinfo{author}{\bibfnamefont{R.~V.} \bibnamefont{Harlander}}
  \bibnamefont{and} \bibinfo{author}{\bibfnamefont{W.~B.}
  \bibnamefont{Kilgore}}, \bibinfo{journal}{Phys. Rev. Lett.}
  \textbf{\bibinfo{volume}{88}}, \bibinfo{pages}{201801}
  (\bibinfo{year}{2002}), \eprint{hep-ph/0201206}.

\bibitem[{\citenamefont{Anastasiou and Melnikov}(2002)}]{Anastasiou:2002yz}
\bibinfo{author}{\bibfnamefont{C.}~\bibnamefont{Anastasiou}} \bibnamefont{and}
  \bibinfo{author}{\bibfnamefont{K.}~\bibnamefont{Melnikov}},
  \bibinfo{journal}{Nucl. Phys.} \textbf{\bibinfo{volume}{B646}},
  \bibinfo{pages}{220} (\bibinfo{year}{2002}), \eprint{hep-ph/0207004}.

\bibitem[{\citenamefont{Anastasiou et~al.}(2004)\citenamefont{Anastasiou,
  Dixon, Melnikov, and Petriello}}]{Anastasiou:2003ds}
\bibinfo{author}{\bibfnamefont{C.}~\bibnamefont{Anastasiou}},
  \bibinfo{author}{\bibfnamefont{L.~J.} \bibnamefont{Dixon}},
  \bibinfo{author}{\bibfnamefont{K.}~\bibnamefont{Melnikov}}, \bibnamefont{and}
  \bibinfo{author}{\bibfnamefont{F.}~\bibnamefont{Petriello}},
  \bibinfo{journal}{Phys. Rev.} \textbf{\bibinfo{volume}{D69}},
  \bibinfo{pages}{094008} (\bibinfo{year}{2004}), \eprint{hep-ph/0312266}.

\bibitem[{\citenamefont{Sjostrand et~al.}(2008)\citenamefont{Sjostrand, Mrenna,
  and Skands}}]{Sjostrand:2007gs}
\bibinfo{author}{\bibfnamefont{T.}~\bibnamefont{Sjostrand}},
  \bibinfo{author}{\bibfnamefont{S.}~\bibnamefont{Mrenna}}, \bibnamefont{and}
  \bibinfo{author}{\bibfnamefont{P.}~\bibnamefont{Skands}},
  \bibinfo{journal}{Comput. Phys. Commun.} \textbf{\bibinfo{volume}{178}},
  \bibinfo{pages}{852} (\bibinfo{year}{2008}), \eprint{0710.3820}.

\bibitem[{\citenamefont{Bahr et~al.}(2008)}]{Bahr:2008pv}
\bibinfo{author}{\bibfnamefont{M.}~\bibnamefont{Bahr}} \bibnamefont{et~al.}
  (\bibinfo{year}{2008}), \eprint{0803.0883}.

\bibitem[{\citenamefont{Klamke and Zeppenfeld}(2007)}]{Klamke:2007cu}
\bibinfo{author}{\bibfnamefont{G.}~\bibnamefont{Klamke}} \bibnamefont{and}
  \bibinfo{author}{\bibfnamefont{D.}~\bibnamefont{Zeppenfeld}}
  (\bibinfo{year}{2007}), \eprint{hep-ph/0703202}.

\bibitem[{\citenamefont{Hankele et~al.}(2006)\citenamefont{Hankele, Klamke, and
  Zeppenfeld}}]{Hankele:2006ja}
\bibinfo{author}{\bibfnamefont{V.}~\bibnamefont{Hankele}},
  \bibinfo{author}{\bibfnamefont{G.}~\bibnamefont{Klamke}}, \bibnamefont{and}
  \bibinfo{author}{\bibfnamefont{D.}~\bibnamefont{Zeppenfeld}}
  (\bibinfo{year}{2006}), \eprint{hep-ph/0605117}.

\bibitem[{\citenamefont{Cahn and Dawson}(1984)}]{Cahn:1983ip}
\bibinfo{author}{\bibfnamefont{R.~N.} \bibnamefont{Cahn}} \bibnamefont{and}
  \bibinfo{author}{\bibfnamefont{S.}~\bibnamefont{Dawson}},
  \bibinfo{journal}{Phys. Lett.} \textbf{\bibinfo{volume}{B136}},
  \bibinfo{pages}{196} (\bibinfo{year}{1984}).

\bibitem[{\citenamefont{Dicus and Willenbrock}(1985)}]{Dicus:1985zg}
\bibinfo{author}{\bibfnamefont{D.~A.} \bibnamefont{Dicus}} \bibnamefont{and}
  \bibinfo{author}{\bibfnamefont{S.~S.~D.} \bibnamefont{Willenbrock}},
  \bibinfo{journal}{Phys. Rev.} \textbf{\bibinfo{volume}{D32}},
  \bibinfo{pages}{1642} (\bibinfo{year}{1985}).

\bibitem[{\citenamefont{Altarelli et~al.}(1987)\citenamefont{Altarelli, Mele,
  and Pitolli}}]{Altarelli:1987ue}
\bibinfo{author}{\bibfnamefont{G.}~\bibnamefont{Altarelli}},
  \bibinfo{author}{\bibfnamefont{B.}~\bibnamefont{Mele}}, \bibnamefont{and}
  \bibinfo{author}{\bibfnamefont{F.}~\bibnamefont{Pitolli}},
  \bibinfo{journal}{Nucl. Phys.} \textbf{\bibinfo{volume}{B287}},
  \bibinfo{pages}{205} (\bibinfo{year}{1987}).

\bibitem[{\citenamefont{Plehn et~al.}(2002)\citenamefont{Plehn, Rainwater, and
  Zeppenfeld}}]{Plehn:2001nj}
\bibinfo{author}{\bibfnamefont{T.}~\bibnamefont{Plehn}},
  \bibinfo{author}{\bibfnamefont{D.~L.} \bibnamefont{Rainwater}},
  \bibnamefont{and}
  \bibinfo{author}{\bibfnamefont{D.}~\bibnamefont{Zeppenfeld}},
  \bibinfo{journal}{Phys. Rev. Lett.} \textbf{\bibinfo{volume}{88}},
  \bibinfo{pages}{051801} (\bibinfo{year}{2002}), \eprint{hep-ph/0105325}.

\bibitem[{\citenamefont{Del~Duca et~al.}(2001)\citenamefont{Del~Duca, Kilgore,
  Oleari, Schmidt, and Zeppenfeld}}]{DelDuca:2001fn}
\bibinfo{author}{\bibfnamefont{V.}~\bibnamefont{Del~Duca}},
  \bibinfo{author}{\bibfnamefont{W.}~\bibnamefont{Kilgore}},
  \bibinfo{author}{\bibfnamefont{C.}~\bibnamefont{Oleari}},
  \bibinfo{author}{\bibfnamefont{C.}~\bibnamefont{Schmidt}}, \bibnamefont{and}
  \bibinfo{author}{\bibfnamefont{D.}~\bibnamefont{Zeppenfeld}},
  \bibinfo{journal}{Nucl. Phys.} \textbf{\bibinfo{volume}{B616}},
  \bibinfo{pages}{367} (\bibinfo{year}{2001}), \eprint{hep-ph/0108030}.

\bibitem[{\citenamefont{Campbell et~al.}(2006)\citenamefont{Campbell, Ellis,
  and Zanderighi}}]{Campbell:2006xx}
\bibinfo{author}{\bibfnamefont{J.~M.} \bibnamefont{Campbell}},
  \bibinfo{author}{\bibfnamefont{R.~K.} \bibnamefont{Ellis}}, \bibnamefont{and}
  \bibinfo{author}{\bibfnamefont{G.}~\bibnamefont{Zanderighi}},
  \bibinfo{journal}{JHEP} \textbf{\bibinfo{volume}{10}}, \bibinfo{pages}{028}
  (\bibinfo{year}{2006}), \eprint{hep-ph/0608194}.

\bibitem[{\citenamefont{Del~Duca et~al.}(2006)}]{DelDuca:2006hk}
\bibinfo{author}{\bibfnamefont{V.}~\bibnamefont{Del~Duca}}
  \bibnamefont{et~al.}, \bibinfo{journal}{JHEP} \textbf{\bibinfo{volume}{10}},
  \bibinfo{pages}{016} (\bibinfo{year}{2006}), \eprint{hep-ph/0608158}.

\bibitem[{\citenamefont{Wilczek}(1977)}]{Wilczek:1977zn}
\bibinfo{author}{\bibfnamefont{F.}~\bibnamefont{Wilczek}},
  \bibinfo{journal}{Phys. Rev. Lett.} \textbf{\bibinfo{volume}{39}},
  \bibinfo{pages}{1304} (\bibinfo{year}{1977}).

\bibitem[{\citenamefont{Dawson}(1991)}]{Dawson:1990zj}
\bibinfo{author}{\bibfnamefont{S.}~\bibnamefont{Dawson}},
  \bibinfo{journal}{Nucl. Phys.} \textbf{\bibinfo{volume}{B359}},
  \bibinfo{pages}{283} (\bibinfo{year}{1991}).

\bibitem[{\citenamefont{Djouadi et~al.}(1991)\citenamefont{Djouadi, Spira, and
  Zerwas}}]{Djouadi:1991tk}
\bibinfo{author}{\bibfnamefont{A.}~\bibnamefont{Djouadi}},
  \bibinfo{author}{\bibfnamefont{M.}~\bibnamefont{Spira}}, \bibnamefont{and}
  \bibinfo{author}{\bibfnamefont{P.~M.} \bibnamefont{Zerwas}},
  \bibinfo{journal}{Phys. Lett.} \textbf{\bibinfo{volume}{B264}},
  \bibinfo{pages}{440} (\bibinfo{year}{1991}).

\bibitem[{\citenamefont{Dawson and Kauffman}(1994)}]{Dawson:1993qf}
\bibinfo{author}{\bibfnamefont{S.}~\bibnamefont{Dawson}} \bibnamefont{and}
  \bibinfo{author}{\bibfnamefont{R.}~\bibnamefont{Kauffman}},
  \bibinfo{journal}{Phys. Rev.} \textbf{\bibinfo{volume}{D49}},
  \bibinfo{pages}{2298} (\bibinfo{year}{1994}), \eprint{hep-ph/9310281}.

\bibitem[{\citenamefont{Butterworth et~al.}(2003)\citenamefont{Butterworth,
  Couchman, Cox, and Waugh}}]{Butterworth:2002xg}
\bibinfo{author}{\bibfnamefont{J.~M.} \bibnamefont{Butterworth}},
  \bibinfo{author}{\bibfnamefont{J.~P.} \bibnamefont{Couchman}},
  \bibinfo{author}{\bibfnamefont{B.~E.} \bibnamefont{Cox}}, \bibnamefont{and}
  \bibinfo{author}{\bibfnamefont{B.~M.} \bibnamefont{Waugh}},
  \bibinfo{journal}{Comput. Phys. Commun.} \textbf{\bibinfo{volume}{153}},
  \bibinfo{pages}{85} (\bibinfo{year}{2003}), \eprint{hep-ph/0210022}.

\bibitem[{\citenamefont{Martin et~al.}(2004)\citenamefont{Martin, Roberts,
  Stirling, and Thorne}}]{Martin:2004ir}
\bibinfo{author}{\bibfnamefont{A.~D.} \bibnamefont{Martin}},
  \bibinfo{author}{\bibfnamefont{R.~G.} \bibnamefont{Roberts}},
  \bibinfo{author}{\bibfnamefont{W.~J.} \bibnamefont{Stirling}},
  \bibnamefont{and} \bibinfo{author}{\bibfnamefont{R.~S.}
  \bibnamefont{Thorne}}, \bibinfo{journal}{Phys. Lett.}
  \textbf{\bibinfo{volume}{B604}}, \bibinfo{pages}{61} (\bibinfo{year}{2004}),
  \eprint{hep-ph/0410230}.

\bibitem[{\citenamefont{Alwall et~al.}(2007)}]{Alwall:2007st}
\bibinfo{author}{\bibfnamefont{J.}~\bibnamefont{Alwall}} \bibnamefont{et~al.},
  \bibinfo{journal}{JHEP} \textbf{\bibinfo{volume}{09}}, \bibinfo{pages}{028}
  (\bibinfo{year}{2007}), \eprint{arXiv:0706.2334 [hep-ph]}.

\bibitem[{\citenamefont{Del~Duca et~al.}(2004)\citenamefont{Del~Duca, Frizzo,
  and Maltoni}}]{DelDuca:2004wt}
\bibinfo{author}{\bibfnamefont{V.}~\bibnamefont{Del~Duca}},
  \bibinfo{author}{\bibfnamefont{A.}~\bibnamefont{Frizzo}}, \bibnamefont{and}
  \bibinfo{author}{\bibfnamefont{F.}~\bibnamefont{Maltoni}},
  \bibinfo{journal}{JHEP} \textbf{\bibinfo{volume}{05}}, \bibinfo{pages}{064}
  (\bibinfo{year}{2004}), \eprint{hep-ph/0404013}.

\bibitem[{\citenamefont{Fadin et~al.}(1975)\citenamefont{Fadin, Kuraev, and
  Lipatov}}]{Fadin:1975cb}
\bibinfo{author}{\bibfnamefont{V.~S.} \bibnamefont{Fadin}},
  \bibinfo{author}{\bibfnamefont{E.~A.} \bibnamefont{Kuraev}},
  \bibnamefont{and} \bibinfo{author}{\bibfnamefont{L.~N.}
  \bibnamefont{Lipatov}}, \bibinfo{journal}{Phys. Lett.}
  \textbf{\bibinfo{volume}{B60}}, \bibinfo{pages}{50} (\bibinfo{year}{1975}).

\bibitem[{\citenamefont{Del~Duca et~al.}(2003)\citenamefont{Del~Duca, Kilgore,
  Oleari, Schmidt, and Zeppenfeld}}]{DelDuca:2003ba}
\bibinfo{author}{\bibfnamefont{V.}~\bibnamefont{Del~Duca}},
  \bibinfo{author}{\bibfnamefont{W.}~\bibnamefont{Kilgore}},
  \bibinfo{author}{\bibfnamefont{C.}~\bibnamefont{Oleari}},
  \bibinfo{author}{\bibfnamefont{C.~R.} \bibnamefont{Schmidt}},
  \bibnamefont{and}
  \bibinfo{author}{\bibfnamefont{D.}~\bibnamefont{Zeppenfeld}},
  \bibinfo{journal}{Phys. Rev.} \textbf{\bibinfo{volume}{D67}},
  \bibinfo{pages}{073003} (\bibinfo{year}{2003}), \eprint{hep-ph/0301013}.

\bibitem[{\citenamefont{Fadin}(1998)}]{Fadin:1998sh}
\bibinfo{author}{\bibfnamefont{V.~S.} \bibnamefont{Fadin}}
  (\bibinfo{year}{1998}), \eprint{hep-ph/9807528}.

\bibitem[{\citenamefont{Balitsky and Lipatov}(1978)}]{Balitsky:1978ic}
\bibinfo{author}{\bibfnamefont{I.~I.} \bibnamefont{Balitsky}} \bibnamefont{and}
  \bibinfo{author}{\bibfnamefont{L.~N.} \bibnamefont{Lipatov}},
  \bibinfo{journal}{Sov. J. Nucl. Phys.} \textbf{\bibinfo{volume}{28}},
  \bibinfo{pages}{822} (\bibinfo{year}{1978}).

\bibitem[{\citenamefont{Andersen}(2006)}]{Andersen:2006sp}
\bibinfo{author}{\bibfnamefont{J.~R.} \bibnamefont{Andersen}},
  \bibinfo{journal}{Phys. Lett.} \textbf{\bibinfo{volume}{B639}},
  \bibinfo{pages}{290} (\bibinfo{year}{2006}), \eprint{hep-ph/0602182}.

\bibitem[{\citenamefont{Forshaw and Sjodahl}(2007)}]{Forshaw:2007vb}
\bibinfo{author}{\bibfnamefont{J.~R.} \bibnamefont{Forshaw}} \bibnamefont{and}
  \bibinfo{author}{\bibfnamefont{M.}~\bibnamefont{Sjodahl}},
  \bibinfo{journal}{JHEP} \textbf{\bibinfo{volume}{09}}, \bibinfo{pages}{119}
  (\bibinfo{year}{2007}), \eprint{0705.1504}.

\end{thebibliography}
\end{document}